\documentclass{elsart}
\usepackage{epsfig}
\begin{document}
\begin{frontmatter}

\title{Optical pulse propagation in fibers with random dispersion}
\author{F.Kh. Abdullaev\corauthref{cor1}},
\author{D.V. Navotny}
\address{Physical-Technical Institute of the Uzbek Academy of Sciences, \\
2-b, G. Mavlyanov str., 700084, Tashkent, Uzbekistan }

\author{B.B. Baizakov\thanksref{label2}}
\address{Dipartimento di Fisica "E.R. Caianiello" \\
and Istituto Nazionale di Fisica della Materia (INFM), \\
Universit\'a di Salerno, I-84081 Baronissi (SA), Italy}

\corauth[cor1]{Corresponding author. E-mail: fatkh@physic.uzsci.net}
\thanks[label2]{On leave from Physical-Technical Institute,
Tashkent, Uzbekistan.}

\begin{abstract}
The propagation of optical pulses in two types of fibers with
randomly varying dispersion is investigated. The first type refers
to a uniform fiber dispersion superimposed by random modulations
with a zero mean. The second type is the dispersion-managed fiber
line with fluctuating parameters of the dispersion map.
Application of the mean field method leads to the nonlinear
Schr\"odinger equation (NLSE) with a dissipation term, expressed
by a 4th order derivative of the wave envelope. The prediction of
the mean field approach regarding the decay rate of a soliton is
compared with that of the perturbation theory based on the
Inverse Scattering Transform (IST). A good agreement between
these two approaches is found. Possible ways of compensation of
the radiative decay of solitons using the linear and nonlinear
amplification are explored. Corresponding mean field equation
coincides with the complex Swift-Hohenberg equation. The
condition for the autosolitonic regime in propagation of optical
pulses along a fiber line with fluctuating dispersion is derived
and the existence of autosoliton (dissipative soliton ) is
confirmed by direct numerical simulation of the stochastic NLSE.
The dynamics of solitons in optical communication systems with
random dispersion-management is further studied applying the
variational principle to the mean field NLSE, which results in a
system of ODE's for soliton parameters. Extensive numerical
simulations of the stochastic NLSE, mean field equation and
corresponding set of ODE's are performed to verify the
predictions of the developed theory.
\end{abstract}

\end{frontmatter}

\section{Introduction}
Propagation of optical pulses in fibers with non-uniform
dispersion remains to be a field of intensive research. The
relevant studies are motivated by their application to modern
optical communication systems. Random fluctuations of the
dispersion coefficient is inherent to existing optical fibers,
which deteriorates the performance of communication lines based
on both standard and dispersion-managed solitons
\cite{Mollenauer}. Therefore, the proper compensation of the
pulse distortion caused by random fluctuations of the fiber
dispersion is of vital importance, especially for long haul
optical communication systems.

The previous investigations were concerned with the analysis of
the modulational instability (MI) of electromagnetic waves in
fibers with random dispersion. It was shown that the randomness
can reduce the MI gain for anomalous dispersion region and
generate the instability for the whole spectrum of modulations in
the normal dispersion region \cite{Karlsson,Darmanyan,Abd1}.
Recently the performance of dispersion-managed soliton system with
random variations of span length and span path-average dispersions
has been studied numerically in Ref. \cite{xie}.
The comprehensive analysis of the dynamics of solitons implies the
solution of the  nonlinear Schr\"odinger equation (NLSE) with
randomly varying coefficients, which is a rather complex problem.
However, in the particular case, when a uniform fiber
dispersion is superimposed by weak random modulations, the
perturbation theory based on the Inverse Scattering Transform
(IST) can be applied. In the framework of this approach the decay
law for optical solitons was derived which describes their
radiative damping due to emission of linear waves
\cite{Abd2,Abd7}. The case of dispersion-managed (DM) solitons is
more complicated because here one has the superposition of the
random fluctuations and strong periodic modulations of the
dispersion. In recent works \cite{Abd3,Malomed,Bronski,Schafer} the variational
approach was employed to analyze the dynamics of optical pulses in
systems with random dispersion-management. It was shown that the
DM soliton is subject to disintegration due to random variations of the
dispersion map parameters. The averaged equation in the frequency domain
was derived in \cite{Chertkov,Chertkov1,Chertkov2}, where an interesting idea of
pinning (compensation of the accumulated effect of fluctuations) is
proposed. Numerical simulations of the dynamics of DM solitons in
systems with random dispersion-management are presented in
\cite{Abd3,Malomed,Haus,Blow}. Analytical description of the
DM soliton dynamics in fluctuating media remains to be an open
problem, since the radiative effects cannot be taken into account
in the frame of the variational approach. However, there exists a
limit pertaining to strong dispersion-management, where the NLSE
with periodic coefficients is nearly integrable \cite{Zakharov}.
The development of the stochastic perturbation theory similar to
the IST approach may be successful in this limit. This problem
deserves a separate consideration.

In this study we apply the mean field method (MFM) for the analysis of
propagation of optical pulses in fibers with random dispersion, including
the DM case. The advantage of this method consists in that, it doesn't
depend on the fact whether the basic deterministic equation is integrable
or not.

The paper is organized as follows. In section II we derive the
mean field equation for the case of uniform dispersion perturbed
by random modulations and compare the predictions of this approach
with the corresponding result of the IST based perturbation
theory. The condition for the existence of autosoliton in random
media is derived. The section III is devoted to analysis of the DM
soliton dynamics in the framework of MFM. We show that the
variational approach  applied to the mean field equation with a
frequency dependent damping leads to a system of ODE's for DM
soliton parameters. And in the last section IV, we briefly
summarize the main results of this study.

\section{Mean field method and IST approach}

The propagation of optical pulses in a fiber with uniform
dispersion superimposed by random modulations was considered in
the framework of the perturbation theory based on the IST
\cite{Abd2}. Assuming the perturbation to be of the form $R =
\epsilon(z)u_{tt},$ where $\epsilon(z)$ is a weak random process,
the radiative decay of a soliton was calculated. The decay law for
the pulse amplitude was derived from the energy conservation and
the dynamic balance between the radiative component and localized
mode. It was shown that the amplitude of the soliton decays with
the propagation distance as $1/z^{1/4}$ for $z<1/\sigma^2$, where
$\sigma^2$ is the noise level. In addition, the decay rate was
revealed to be highly dependent on the pulse duration $\sim
t_0^{-4}$, i.e. the influence of the randomness is superior on
short pulses, particularly in the femtosecond range.

The extension of this approach for more general cases (e.g.
dispersion-managed solitons) is in general not possible. Therefore,
it is of interest to apply another approach which is independent
on the fact whether the underlying deterministic equation is
integrable or not. One of such possibilities is provided by the
mean field method \cite{Klyatzkin}. Despite the well known
troubles of this approach for random nonlinear wave equations
\cite{Konotop,Abdullaev}, one can obtain reasonably well
description in the case of weak fluctuations, and/or small
distances of propagation.

In this section we apply the MFM and the perturbation theory based on
the IST to the problem of optical pulse propagation in fibers with random
dispersion. By comparing the predictions of these two approaches we
reveal the limits of validity of the MFM.

Optical pulse propagation in fibers is well described by the NLSE
\begin{equation} \label{main}
iu_z+\frac{d(z)}{2}u_{tt}+|u|^2u = i\delta u + i\mu |u|^2 u.
\end{equation}
Here $\delta, \ \mu$ are the coefficients of linear and nonlinear
amplification (damping) respectively. In our model these terms are
supposed to compensate the damping originated from the randomness of
the fiber dispersion.
Below we consider the case, when the dispersion coefficient $d(z)$ is the
sum of the constant and random parts,  i.e. $d(z) = 1 + \epsilon(z)$ with
\begin{equation} \label{noise}
<\epsilon> =0, \ <\epsilon(z)\epsilon(z')> =B(z-z',l_{c})_{l_{c}\rightarrow 0}
\rightarrow \sigma^2 \delta(z-z').
\end{equation}
The noise is assumed to be small compared to the constant part of the
dispersion.

To derive the equation averaged over fluctuations we apply the
mean field method, representing the field as consisting of the
mean value and small fluctuating part $u = <u> + \delta u,
<\delta u> = 0, \delta u \ll <u> $, where $<u>$ is a slowly
varying mean field. According to this method we can use the
decoupling $<|u|^2 u> \approx |<u>|^2 u$. This corresponds to
neglecting by fluctuations of the nonlinearity, and the
approximation is valid for propagation distances $z <<
1/\sigma^{2}$. In typical optical experiments this requirement is
satisfied. In general the decoupling procedure is inaccurate
since the scattered field $\delta u$ grows with propagation
distance and becomes comparable with $<u>$, that violates the
assumption $\delta u \ll <u>$. In order to decompose the mean
$<\epsilon(z)u_{tt}>$ we use the Furutsu-Novikov formula
\begin{equation} \label{furnov}
  <\epsilon(z)F(u)> = \int_{0}^{z}B(z-z')<\frac{\delta F(u)}
  {\delta\epsilon(z')}>dz'.
\end{equation}
Of particular interest is the case of white noise fluctuations, when
$B(x)= \sigma^{2}\delta(x)$. Also the causality principle will be useful
\begin{equation}
\frac{\delta u(z)}{\delta \epsilon(z^{\prime})} = 0, \quad
\mbox{ if } \ z^{\prime} < 0, \ \ z^{\prime} > z.
\end{equation}
Then integrating Eq.(\ref{main}) over $z'$ from $0$ to $z$ and taking
variational derivative (with the causality principle in mind) we obtain
\begin{equation}
<\epsilon(z)u_{tt}(z,t)> = \frac{i\sigma^{2}}{4}<u>_{tttt}.
\end{equation}
In the result we get the equation for averaged pulse profile with a
fourth order dissipative term (for simplicity we dropped averaging
symbol in $<u(z,t)>$)
\begin{equation} \label{u4t}
iu_z+\frac{1}{2}u_{tt}+|u|^2u = -i\gamma u_{tttt}+i\delta u+i\mu |u|^2 u,
\end{equation}
where $\gamma = \sigma^2/4, \ \sigma$ being the noise strength.

As it is apparent from this equation, the effect of random
dispersion on the pulse evolution is described as the pulse
propagation in a uniform medium with {\it the frequency depending
dissipation}. Formally this equation coincides with the complex
Swift-Hohenberg equation. Thus we can expect the existence of
{\it dissipative solitons} in nonlinear media with fluctuating dispersion.

At first we examine the decay rate of a pulse under the action of
the 4th order dispersion coefficient in Eq.(\ref{u4t}).
For the weak noise case this term is small and we can apply the perturbation
theory. The single soliton propagation is conveniently described using
the equations for the energy $N = \int|u|^2 dt$ and momentum
$P = \mbox{Im}(\int u_t^* u dt)$.
\begin{equation} \label{dndz}
  \frac{dN}{dz} = 2\int_{-\infty}^{\infty}[\delta |u|^2 + \mu |u|^4 -
  \gamma |u_{tt}|^2]dt,
\end{equation}
\begin{equation} \label{dpdz}
  \frac{dP}{dz}= 2\mbox{Im}\int_{-\infty}^{\infty}[\delta u + \mu |u|^2 u
  + \gamma u_{tttt}]u_t^* dt.
\end{equation}
We look for the single soliton solution in the form
\begin{equation} \label{sech}
  u(z,t) = \eta \mbox{sech}[\eta (t + \Omega z)]e^{-i\Omega t +
  i(\eta^2 - \Omega^2 )z/2}.
\end{equation}
Substituting Eq.(\ref{sech}) into Eqs. (\ref{dndz}),(\ref{dpdz})
we obtain the system of equations for the soliton amplitude and velocity
\begin{eqnarray}
  \frac{d\eta}{dz} &=& 2\eta[\delta - \gamma (\frac{7}{15}\eta^4 + 2\eta^2
  \Omega^2 + \Omega^4 ) + \frac{2}{3}\mu \eta^2 ], \label{ampl} \\
  \frac{d\Omega}{dz}&=&-\frac{8}{3}\gamma\Omega\eta^2
  (\frac{7}{5}\eta^2+\Omega^2).  \label{vel}
\end{eqnarray}
Parameters $\delta, \mu $ are absent in the second equation, since the
soliton velocity is not affected by such type of perturbations.
For $\Omega = 0$ one can find the decay law of the soliton due to
the 4th order dispersive dissipation, which follows from Eq.(\ref{ampl})
\begin{equation}\label{mf}
  \eta (z) = \frac{\eta_0}{(1 + \frac{14}{15}\sigma^2\eta_0^4 z)^{1/4}}
  \approx \eta_0 (1 - \frac{7}{30}\sigma^2 \eta_0^4 z + O(\sigma^4 )).
\end{equation}
Now we can compare the prediction of the MFM
with that of the perturbation theory based on the IST
\cite{Abd2} and thus find the limits of applicability of the MFM
(within the limits of the error of the IST method).
In the frame of IST based perturbation approach, the decrease of the soliton
amplitude due to radiation of random waves under fluctuations of dispersion
can be estimated from the balance equation for the
norm $N = \int|u|^2 dt $, which is the integral of motion.
The norm can be represented as
\begin{equation} \label{norm}
N = 2\eta + \frac{1}{\pi}\int^{\infty}_{-\infty}
\mbox{Ln}|a(\lambda)|^{-2}d\lambda,
\end{equation}
where $\lambda = 2k$ is the spectral parameter, $k$ is the wavenumber,
$a(\lambda), b(\lambda)$ are the Jost coefficients of the Zakharov-Shabat
linear spectral problem and $|a|^2 + |b|^2 =1$. For the
weak noise case $\mbox{Ln}(|a|^{-2}) \approx |b(\lambda)|^2$.
The balance equation can be obtained by differentiation of the norm
with respect to $z$
\begin{equation} \label{BE}
  \frac{d<N>}{dz} = -\frac{1}{2\pi}\int_{-\infty}^{\infty}
  Q(\lambda) d\lambda
  - \frac{1}{\pi}\int_{-\infty}^{\infty}(|b|^2)_{\eta}\frac{d\eta}{dz}d\lambda
  + 4\delta \eta + \frac{8}{3}\eta^3,
\end{equation}
where
\begin{displaymath}
  Q(\lambda) = 2{\rm Re}<b(\lambda)\frac{db^{*}(\lambda)}{dz}>,
\end{displaymath}
is the spectral power of emitted radiation by the soliton.
The Jost coefficient $b(\lambda)$ can be calculated using the perturbation
theory based on the IST \cite{Karpman,Kaup}.
It should be noted that the total wave is given by
\begin{equation}
  u(z,t) = u_{s}(z,t) + u_{c}(z,t),
\end{equation}
where
\begin{eqnarray}
  u_{c} = \frac{1}{i\pi}\int_{-\infty}^{\infty}\frac{b}{a}(\lambda)
  \frac{(\lambda + i\eta \tanh(z))^2}{(\lambda +
  i\eta)^2}e^{2i\lambda t}d\lambda + \nonumber\\
  \frac{\nu^2}{i\pi}\frac{\exp(2i(2\xi (t-t_s )+
  \phi_{s}))}{\cosh(z)^2}\int_{-\infty}^{\infty}
  \frac{b^{\ast}}{a^{\ast}}(\lambda)\frac{1}{\lambda -
  i\eta}e^{-2i\lambda t}d\lambda.
\end{eqnarray}
Here $u_{s}$ is the soliton part of the total wave. The first component
of $u_{c}$ represents the emitted wavepacket. The second component
is due to interaction between the soliton and the
emitted wavepacket and it is appreciable only nearby the soliton.
We will be interested in the decay of a soliton under emission far from the
soliton, assuming that this emission is lost and has no back action on the
soliton.

The equation for the amplitude is
\begin{equation}\label{Eq}
  \frac{d\eta}{dz} = -\frac{\epsilon \eta^5}{1 + 5 \epsilon z \eta^4}, \quad
  \epsilon = \frac{4}{15}\sigma^2.
\end{equation}
The solution reduces to the algebraic equation
\begin{equation}\label{Eq1}
  \eta + \epsilon z \eta^5 = \eta_{0}.
\end{equation}
Since $\eta-<\eta> \sim \epsilon$ we can replace
$<\eta^{5}> \approx <\eta >^5$.

For distances of propagation $z < 1/\sigma^2$ one can neglect
the second term in the denominator of Eq.(\ref{Eq}) and find
the decay law for the amplitude
\begin{equation}\label{Eq2}
  <\eta (z)> = \frac{\eta_0}{(1 + \frac{16}{15}\sigma^2
  \eta_0^4 z)^{1/4}} \approx \eta_{0}(1 - \frac{4}{15}\sigma^{2}
  \eta_{0}^{4}z + O(\sigma^4 )).
\end{equation}
This equation gives almost the same decay rate as Eq.(\ref{mf})
(see \cite{note}).

Thus, the IST approach conforms with the picture of soliton
propagation in uniform media with the effective frequency
dependent ($\sim \omega^4$) losses. Also the IST and MFM predict
the strong $\sim 1/t_0^4$ dependence of the decay rate on the
pulse duration. The comparison shows that the mean field theory
underestimates the soliton decay rate relative to the IST as
given by the ratio $\beta = r_{MFM}/r_{IST} = 0.875$. This
discrepancy between numerical constants in Eq.(\ref{mf}) and
Eq.(\ref{Eq2}) is due to the approximate character of the mean
field equation, where we have neglected by the renormalization of
the nonlinear term.

\begin{figure}[htbp]
\centerline{\includegraphics[width=10.0cm,height=8.0cm,clip]{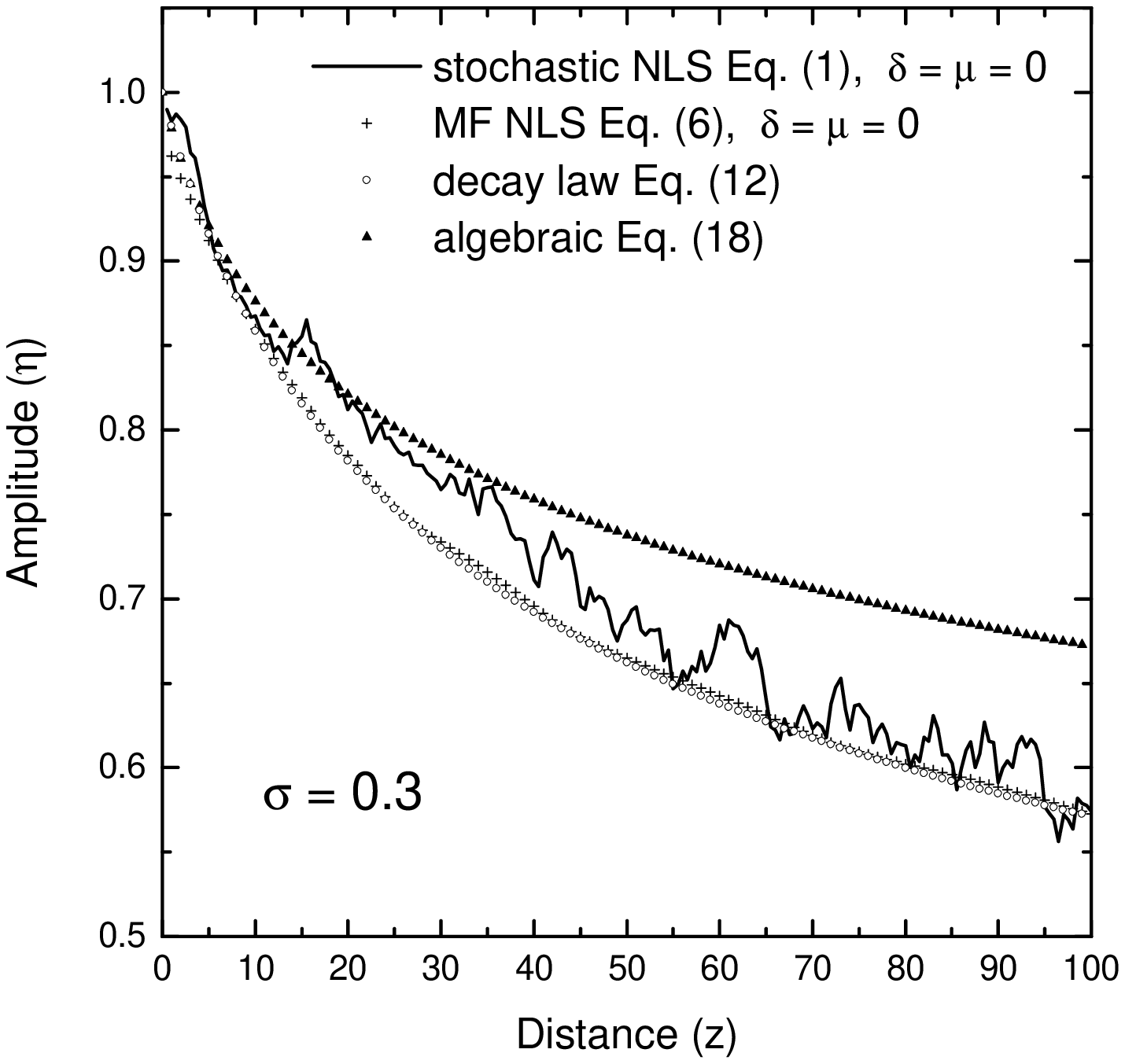}}
\vspace*{0.5cm}
\caption{Decay law for the pulse amplitude according to different models.
The initial pulse has the form $u(0,t)=\eta_0 \mbox{sech}(\eta_0 t)$
with $\eta_0=1$. The fiber dispersion randomly varies around
$d_0=1$ with a mean square fluctuations \  $\sigma=0.3$. }
\label{fig1}
\end{figure}

In Fig.\ref{fig1} we report the comparison of predictions of different models
regarding the decay law of the pulse (decreasing of its amplitude) in the
course of propagation along a fiber line with randomly varying dispersion.
It is seen that all the models well describe the initial stage of the
radiative damping. The MF NLS Eq.(\ref{u4t}) and decay law Eq.(\ref{mf})
better describe the damping at distances $z > 1/\sigma^2$, compared to
the algebraic Eq.(\ref{Eq1}). The curve labeled stochastic NLS is obtained by the averaging over
 70 realizations.

Numerical integration of the stochastic NLS Eq.(\ref{main}) and MF
NLS Eq.(\ref{u4t}) are performed by the split-step fast Fourier
transform. Absorption on the domain boundaries is
employed to imitate the infinite length. The system ODE's
Eq.(\ref{ampl}), Eq.(\ref{vel}) and Eq.(\ref{DM-ODE}) are solved
using the procedure DOPRI8 \cite{Hairer}, which is based on the
Runge-Kutta scheme with adaptive stepsize control.

Inspecting the equation for the field momentum $P$ one can see
that $\Omega_{z} = 0$. This is due to the fact that
the numbers of quanta emitted by the soliton in forward and backward
directions are equal. Therefore the predictions of the MFM and IST for
$\Omega$ coincide only at the fixed point, where
$\Omega = 0$. The ratio of amplitudes at the fixed point is
$r = \eta_{MFM}/\eta_{IST} \approx 0.97$ for the case of linear
amplification, and $r = 0.93$ for nonlinear amplification.

\begin{figure}[htbp]
\centerline{\includegraphics[width=6cm,height=6cm,clip]{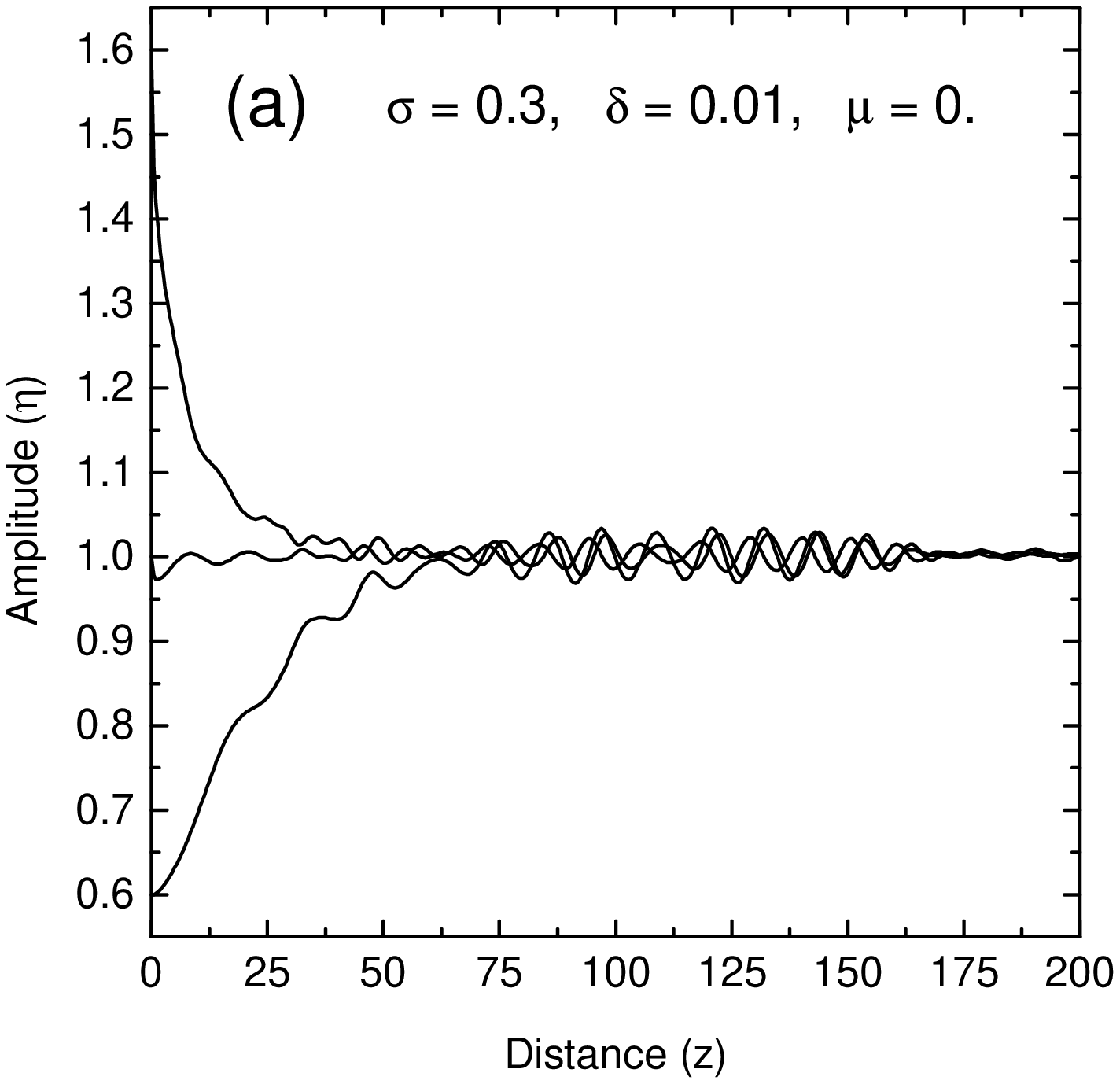} \quad
            \includegraphics[width=6cm,height=6cm,clip]{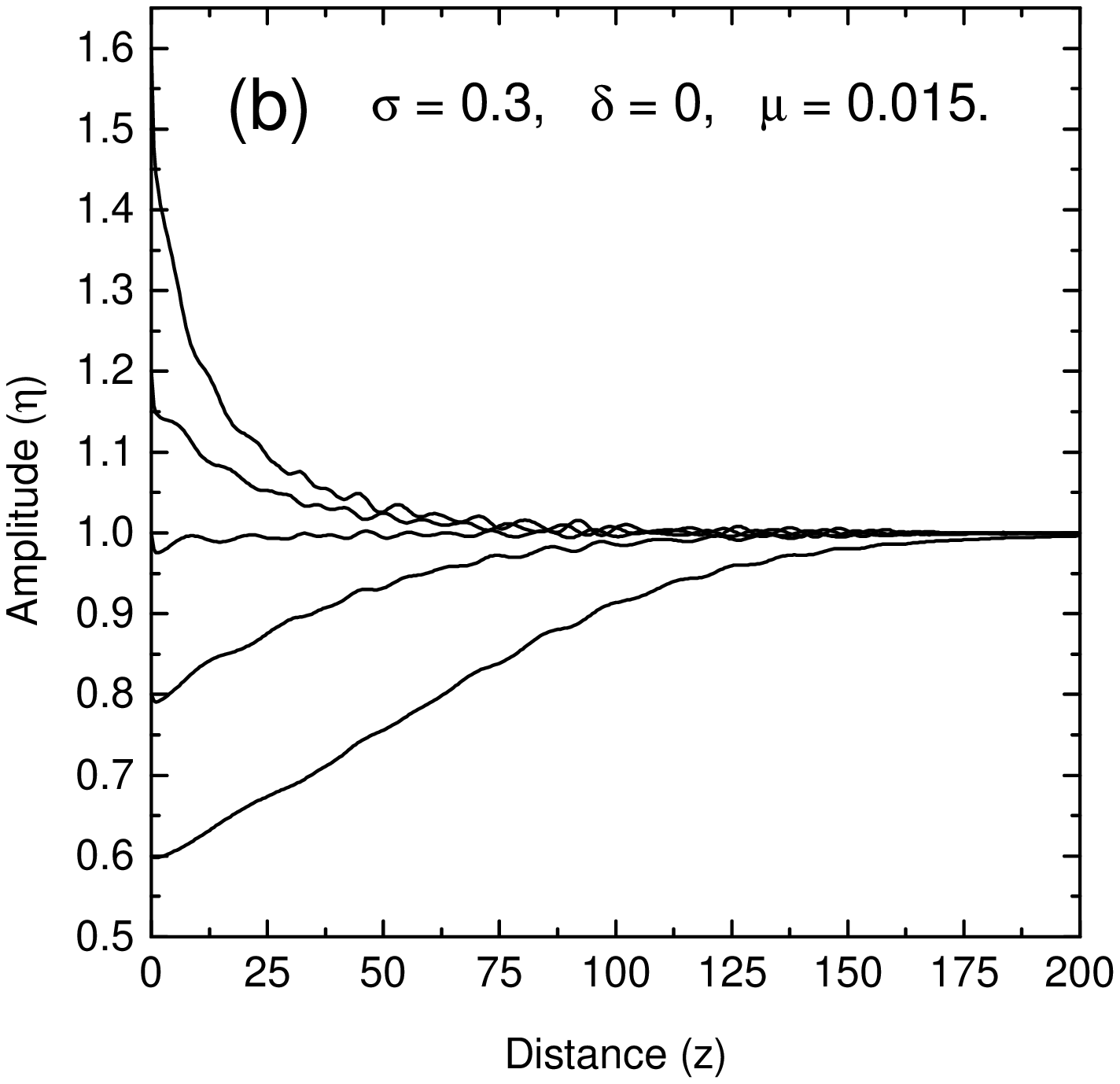}}
\vspace*{0.5cm}
\caption{Autosoliton acquires its stable form at some propagation
distance when its amplitude initially assigned the value
greater or lower than corresponding fixed point value $\eta=1.0$.
Obtained by numerical solution of the MF NLS Eq.(\ref{u4t}) for
$\sigma=0.3, \ \gamma= \sigma^2/4 = 0.0225$, with:
(a) linear amplification $\delta=0.01, \ \mu=0.0$, and
(b) nonlinear amplification $\delta=0.0, \ \mu=0.015.$}
\label{fig2}
\end{figure}

The decay of a soliton can be compensated by linear and/or
nonlinear amplifications. This results in formation of an
autosoliton. In our case, when the pulse power loss due to 4th
order dispersion effect $i\gamma u_{tttt}$ is compensated by the
linear $i\delta u$ and/or nonlinear $i\mu |u|^2 u$ amplification
in Eq.(\ref{main}), such an autosoliton can be created. A
distinctive property of these solitons is that, they recover the
stable waveform when deformed. Fig.\ref{fig2} demonstrates such
an aspect of autosolitons, when the power dissipation due to 4th
order dispersion term is compensated by linear/nonlinear
amplification. Note that the linear amplification gives rise to
more pronounced oscillations around the fixed point during
transient period compared to nonlinear amplification. At longer
propagation distances the oscillations are dumped out and a
stable dissipative soliton is formed.

The amplitude of the autosoliton at the fixed point can be found
from the balance equation (\ref{BE}) of the IST
\begin{equation}
  \eta^{2}_{1,2 st} = \frac{5\mu}{2\sigma^2}\Biggl(1 \pm \sqrt{1 +
  \frac{6\delta \sigma^2}{5\mu^2}}\Biggr).
\end{equation}
Two types of fixed points exist. The first type $\eta_{1st}$ is
defined by the competition between the dissipation (induced by the
randomness of the fiber dispersion) and the linear/nonlinear
amplifications.

For $\delta = 0, \mu >0$ (the nonlinear amplification dominates)
we find
\begin{equation} \label{nonla}
  \eta^{2}_{1st} = \frac{5\mu}{\sigma^2},
\end{equation}
and similarly for $\mu =0, \delta >0$ (linear amplification dominates)
\begin{equation} \label{lina}
  \eta^2_{1st} = \sqrt{\frac{15\delta}{2\sigma^2}}.
\end{equation}
The second type of fixed points is defined by the competition between the
linear dissipation and nonlinear amplification.
\begin{equation} \label{both}
  \eta_{2st} \approx \sqrt{\frac{3|\delta|}{2\mu}} -\frac{9|\delta|^2
  \sigma^2}{20\mu^3},
\end{equation}
i.e. the fluctuations reduce the value of amplitude in this case.

The fixed points for the soliton amplitude (at $\sigma=0.3,
\delta=0.01, \ \mu=0.015$) as predicted by Eqs.(\ref{nonla}),
(\ref{lina}) are as follows : $\eta_{1st}=(5\mu/\sigma^2)^{1/2}
\simeq 0.91$, and $\eta_{1st}=(15\delta/2\sigma^2)^{1/4} \simeq
0.96 $. The fixed points of ODE (\ref{ampl}) for the same
parameters give $\eta=(40 \mu/7 \sigma^2)^{1/2} \simeq 0.976$,
$\eta=(60 \delta/7\sigma^2)^{1/4} \simeq 0.988$, which is the
better approximation to PDE simulation of Fig.\ref{fig2}, when we
assume that the effect of noise $\sigma$ is accounted by the 4th
order dispersive dissipation $\gamma=\sigma^2/4$.

\begin{figure}[htbp]
\centerline{\includegraphics[width=6cm,height=6cm,clip]{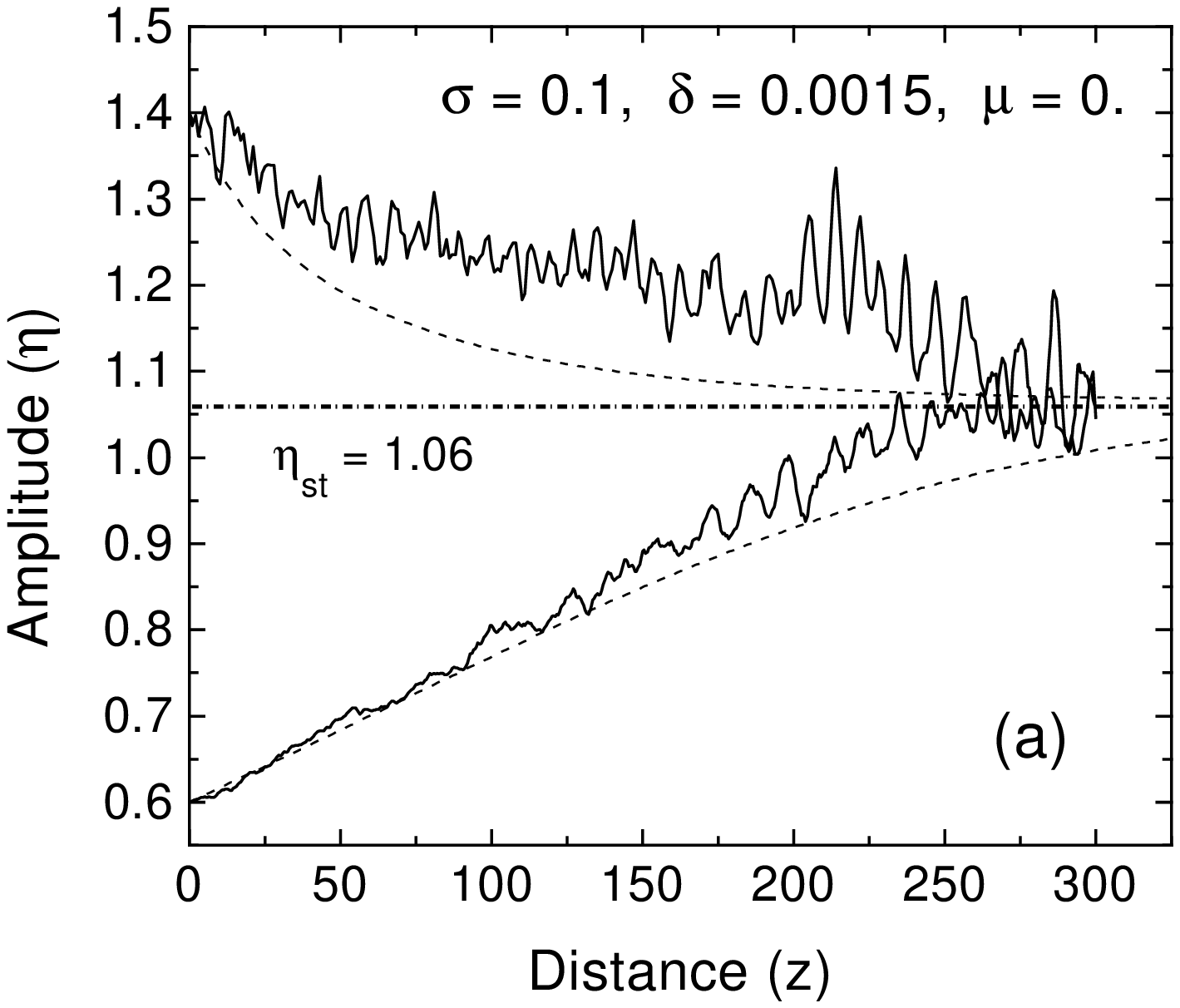} \quad
            \includegraphics[width=6cm,height=6cm,clip]{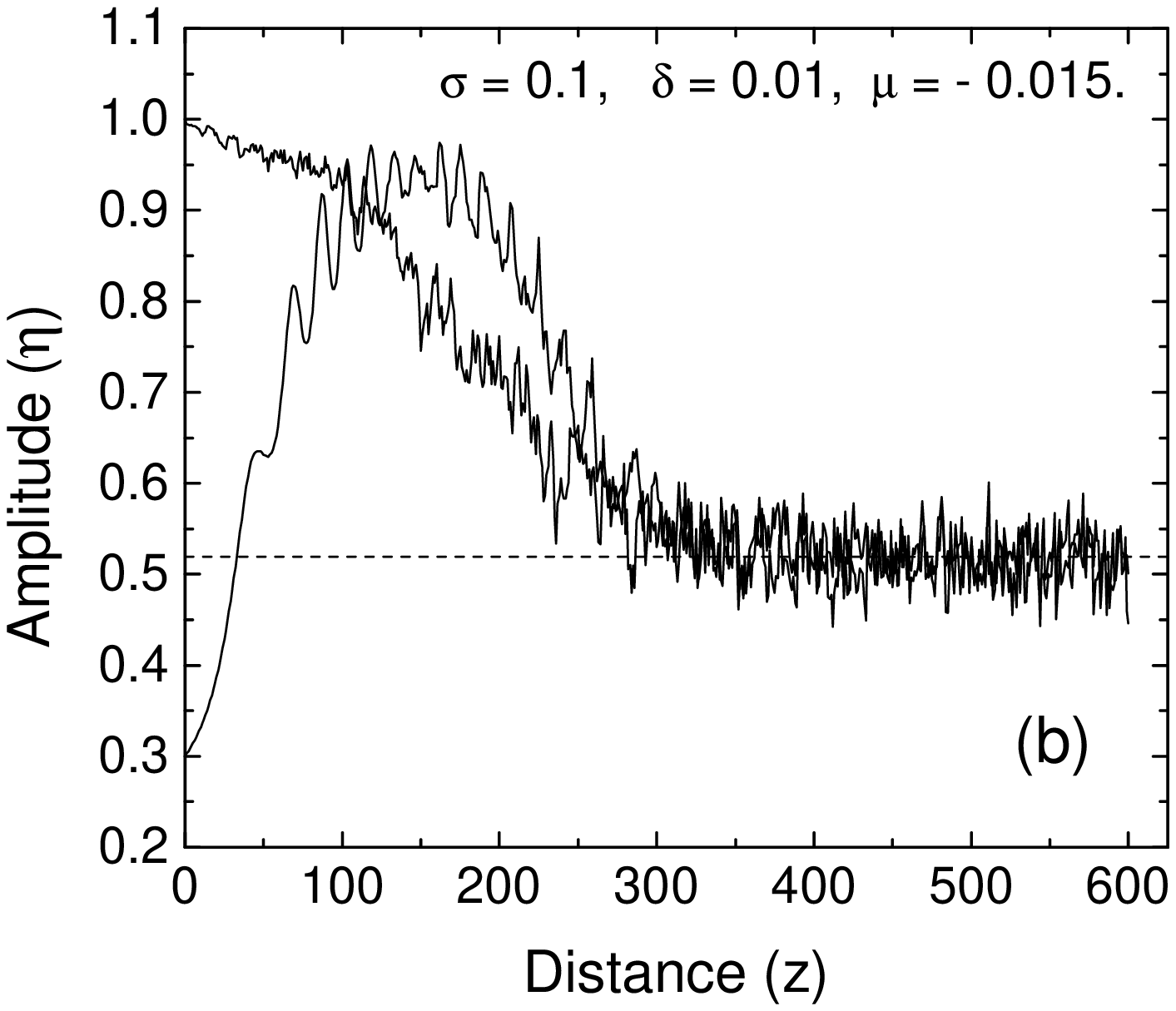}}
\vspace*{0.5cm} \caption{Autosoliton from stochastic NLSE
simulations. (a) Evolution of the soliton's amplitude towards the
stationary value as obtained by numerical solution of the
stochastic Eq.(\ref{main}) (solid lines), and as predicted by the
ODE model Eq.(\ref{ampl}) (dashed lines). Dash-dot line is the
stationary value of the amplitude. (b) Autosoliton resulted from
the combined effect of a linear amplification and nonlinear
dissipation.} \label{fig3}
\end{figure}

Simulations of the stochastic NLS Eq.(\ref{main}), is presented in
Fig.\ref{fig3}. Compensation of the damping (due to the randomness
of the dispersion) by a linear amplification gives rise to a
stable soliton. To verify the property of autolitons observed in
Fig.\ref{fig2}, we assigned the amplitude greater and lower
values, compared to the stationary one $\nu_{st}=1.06$. As
expected, the soliton adjusts itself to the stationary amplitude
at some propagation distance. Similar behavior is observed when
the combination of linear amplification and nonlinear damping is
applied (Fig.\ref{fig3}b). However, it should be pointed out that
the stable pulse propagation is limited by the growth of the zero
mode under amplification. When the initial pulse amplitude is
close to this solution the instability starts to   at
distances $z \sim 1/\mu$ (see f.e.\cite{Akhmediev}).

\begin{figure}[htbp]
\centerline{\includegraphics[width=8cm,height=8cm,clip]{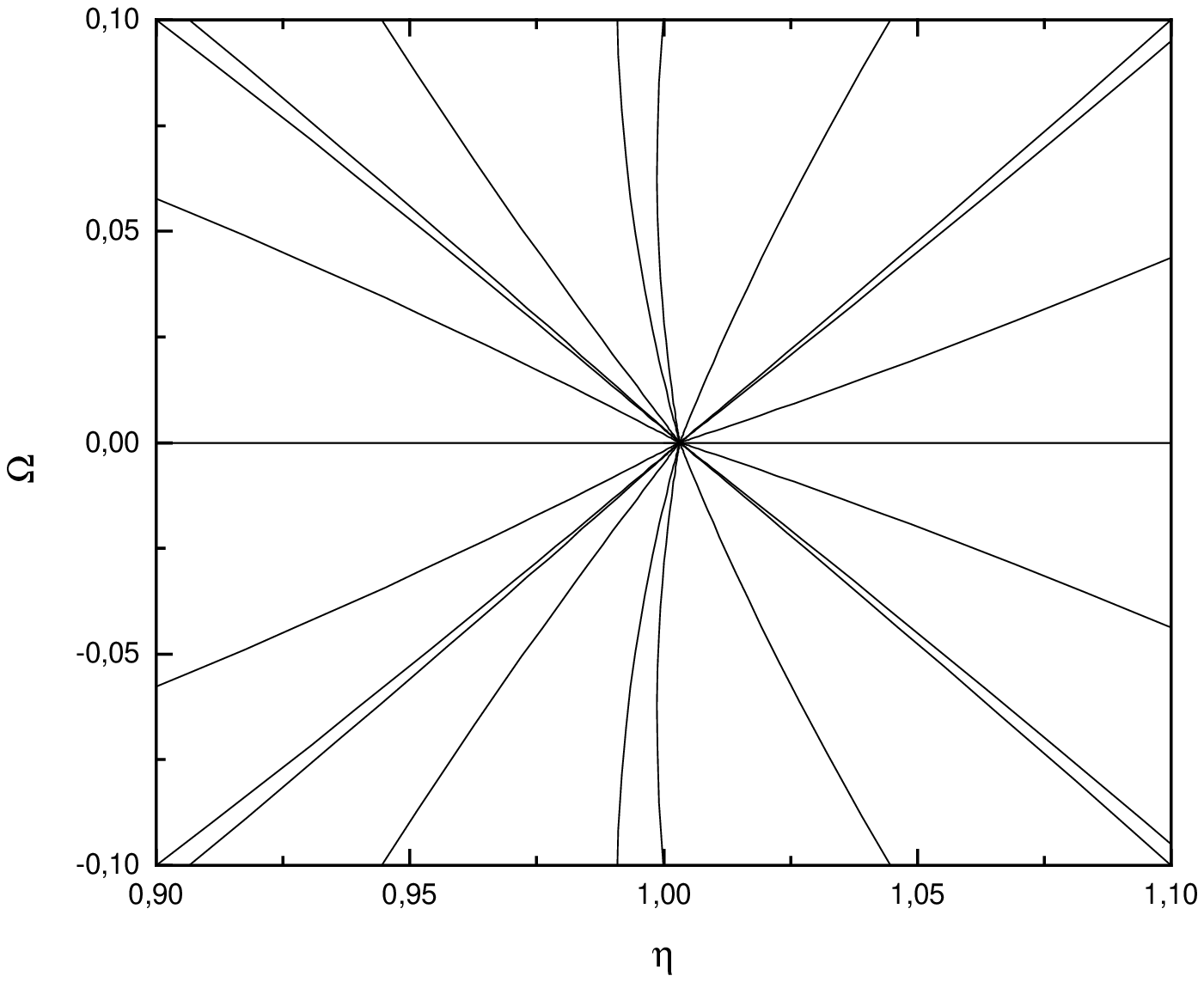}}
\caption{The phase trajectories corresponding to ODE model
Eq.(\ref{ampl}) and Eq.(\ref{vel}). $\sigma=0.1, \ \delta=0.0015,
\ \mu=0.$ } \label{fig4}
\end{figure}

The phase trajectories according to Eq.(\ref{ampl}) and Eq.(\ref{vel})
is shown in Fig.\ref{fig4}. The fixed point appears to be of sink type.

Thus, summarizing the results of this section we note, that the
optical pulse propagation in fibers with random dispersion can be
described as propagation in a uniform fiber with the effective
frequency dependent damping. The effect of randomness of the
fiber dispersion is accounted by a term with 4th order derivative
of the wave envelope. Application of this idea to
dispersion-managed solitons will be considered in the next
section.

\section{Dispersion-managed soliton in a fiber with random dispersion}

The dispersion-managed optical communication line consists of
fiber sections with alternating anomalous and normal dispersion
coefficients \cite{Doran}. Now the function $d(z)$ is the sum of periodically
varying dispersion $d_0(z)$ and the random part $d_1(z)$ i.e $d(z)
= d_{0}(z) + d_{1}(z)$. The usual scale of the DM map is $\sim$
100 km, so we can consider the random modulations of the
dispersion in the white noise limit with $\delta(z)$ correlation.
Due to strong variations of the dispersion within the unit cell
(the strong dispersion-management regime), locally one has the
linear dynamics. The effect of nonlinearity is small in the scale
of a unit cell, while accumulated on longer distances, can lead to
formation of a breathing DM soliton. Therefore we can expect
that the influence of the random modulations of dispersion will
result in the frequency dependent damping in the form calculated
in the previous section and the nonlinear corrections to this
damping will be small.

Performing the averaging over fluctuations yields the following
equation
\begin{equation} \label{u4t-dm}
  iu_z + \frac{d_{0}(z)}{2}u_{tt} + |u|^2 u = -i\gamma u_{tttt} +
  i \delta u + i\mu |u|^2 u.
\end{equation}
In the case of strong dispersion-management
(when $d(z) = (1/\epsilon)d(z/\epsilon), \epsilon \ll 1$,
the averaging procedure can be justified \cite{Zhar}.
Introducing the variable $\zeta = z/\epsilon$ we obtain the equation
\begin{equation}\label{main1}
  iu_{\zeta} + \frac{d(\zeta)}{2}u_{tt} + \epsilon (|u|^2 - i\delta -i\mu |u|^{2})u =0,
\end{equation}
where $d(z) = d_{0}(\zeta) + d_1(\zeta), d_1 \sim
\epsilon^{1/2}$. Let us consider the case when $d_1 \sim
\sqrt{\epsilon}$ and thus  $d_1^2 \sim \epsilon$. One can
decompose the field as $u = <u> + \delta u, \delta u \sim
\sqrt{\epsilon}$, and obtain the following estimates for the
correction $|\delta u|^2 , \delta u^2 \sim \epsilon$. Thus, the
nonlinear corrections from fluctuations to the averaged equation
are of order $\epsilon^2 $ and can be neglected in comparison
with terms $<d_{1}(\zeta)u_{tt}> \sim \epsilon |<u>|^2 <u> \sim
\epsilon$.

The variational approach is proved to be effective for exploring
the dispersion-managed soliton  \cite{Gabitov,Malomed1}. In the
presence of nonconservative terms in Eq.(\ref{u4t-dm}) we can use
the modified variational equations. The equations for the pulse
parameters are \cite{Anderson,Maimistov}

\begin{equation}\label{ve}
  \frac{\partial L}{\partial\eta_i} -
  \frac{d}{dz}\frac{\partial L}{\partial\eta_i} =
  \int_{-\infty}^{\infty}(R\frac{\partial u^*}{\partial\eta_i} + c.c.)
\end{equation}
Here $L$ is the averaged Lagrangian $L = \int L(z,t)dt$.
Employing the Gaussian anzats for the waveform
\begin{equation}\label{pulse}
  u(z,t)=A(z) \exp[-\frac{t^2}{2a(z)^2} + i\frac{b(z) t^2}{2} + i\phi (z)].
\end{equation}
we obtain the following system of ODE's from the Eq.(\ref{ve})
\begin{eqnarray} \label{DM-ODE}
  (a^{2})_{z} &=& 2d_0 (z) a^2 b + 6 \gamma a^2
  [\frac{1}{a^4} - b^4 a^4 ] -\frac{5}{2\sqrt{2}}\mu A^{2}a^{2},\nonumber\\
  (A^2 )_{z} &=& -d_{0} (z) A^2 b - \frac{3}{2}
  \gamma A^2 [\frac{3}{a^4} +  2b^2 - b^4 a^4 ]
  + 2\delta A^{2} + \frac{13}{4\sqrt{2}}\mu A^{4},\nonumber\\
  b_{z} &=& \frac{d_0(z)}{a^4} - d_0 (z) b^2 -
  \frac{A^2}{\sqrt{2}a^2 } - 12 \gamma b (\frac{1}{a^4} + b^2 ).
\end{eqnarray}
When $\gamma = \delta = \mu = 0$, we reproduce the variational
equations for a DM soliton  \cite{Gabitov}.

\begin{figure}[htbp]
\centerline{\includegraphics[width=10cm,height=8cm,clip]{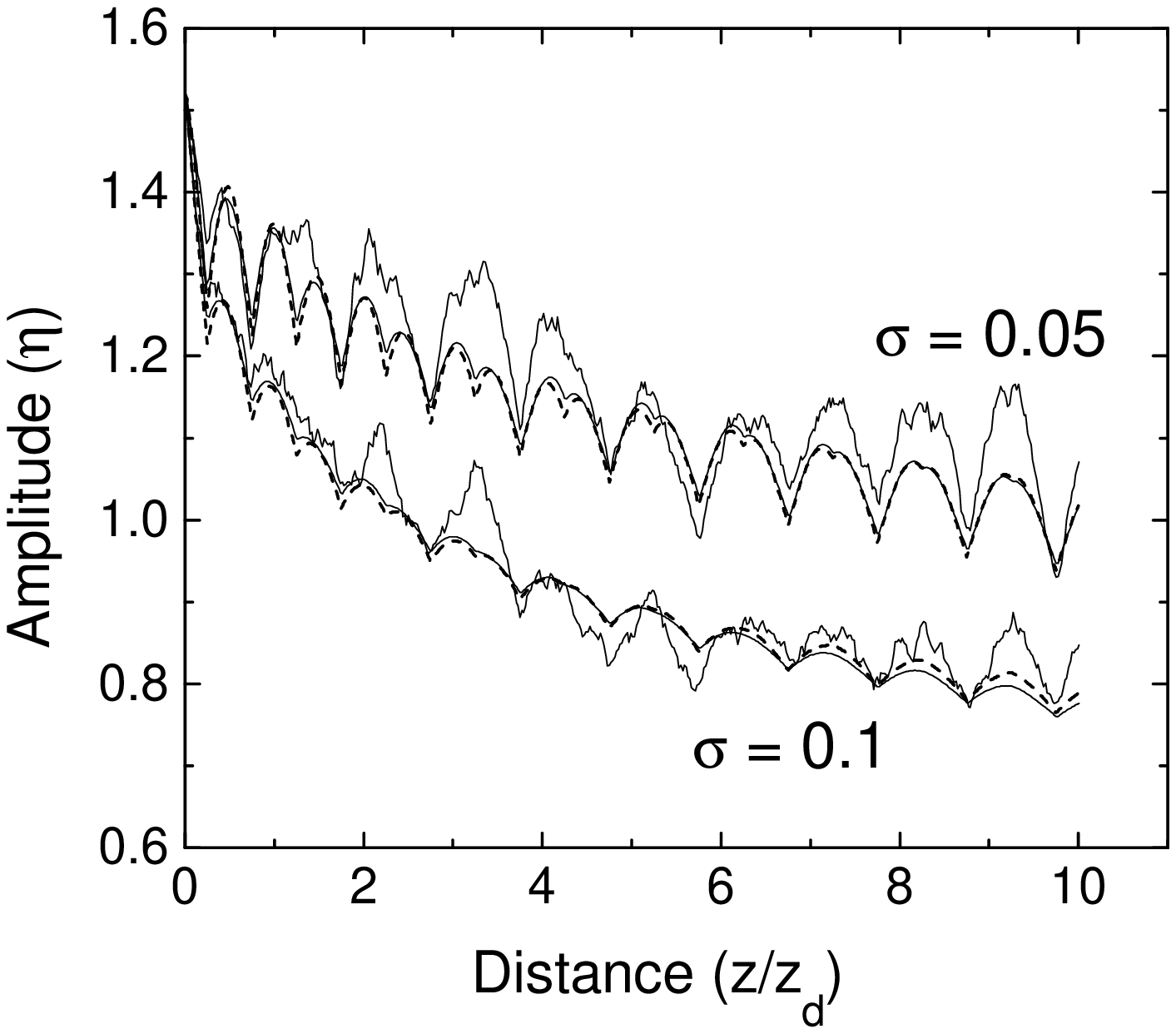}}
\caption{(a) Comparison of pulse decay rates obtained by numerical
integration of the stochastic NLS Eq.(\ref{main}) \ (broken
line), the averaged equation Eq.(\ref{u4t-dm}) \ (solid line), and
system of ODE's  (\ref{DM-ODE}) \ (dashed line). DM map
parameters and initial conditions are $D_1=30, \ D_2=-28, \
z_1=z_2=0.078, \ z_d=0.156, \ A(0)=1.52, \ a(0)=1.059, \ b(0) = 0,
$ and $\gamma = \sigma^2/4, \ \sigma = 0.05 \div 0.1, \ \delta =
\mu = 0.$} \label{fig5}
\end{figure}

In order to verify the validity of the mean field equation
(\ref{u4t-dm}) and corresponding variational equations
(\ref{DM-ODE}) for description of the DM soliton dynamics we
compared the result of their numerical solution with that of the
original stochastic equation (\ref{main}), averaged over 50
realisations of random paths. As an initial condition we take the
waveform Eq.(\ref{pulse}). Parameters of the dispersion map are:
$D_1=30, \ D_2=-28, \ z_1=z_2=0.078,  \ z_d=z_1+z_2$.
Fluctuations of the fiber dispersion coefficient is supposed to
be of type Eq.(\ref{noise}), and $d(z)=d_0(z)[1+\epsilon(z)]$.
Satisfactory agreement of these data is displayed in
Fig.\ref{fig5}.
A characteristic feature of the pulse decay process due to the
4th order dispersion term is that, the pulse amplitude rapidly
decreases at initial stage, after that almost linear law is
followed. This behavior can be understood from the fact, that
$u_{tttt}$ term depends on the  pulse shape: its value is bigger
when the pulse is more sharp in the initial stage, and becomes
small when the pulse is dumped and broadened.

\begin{figure}[htbp]
\centerline{\includegraphics[width=6cm,height=6cm,clip]{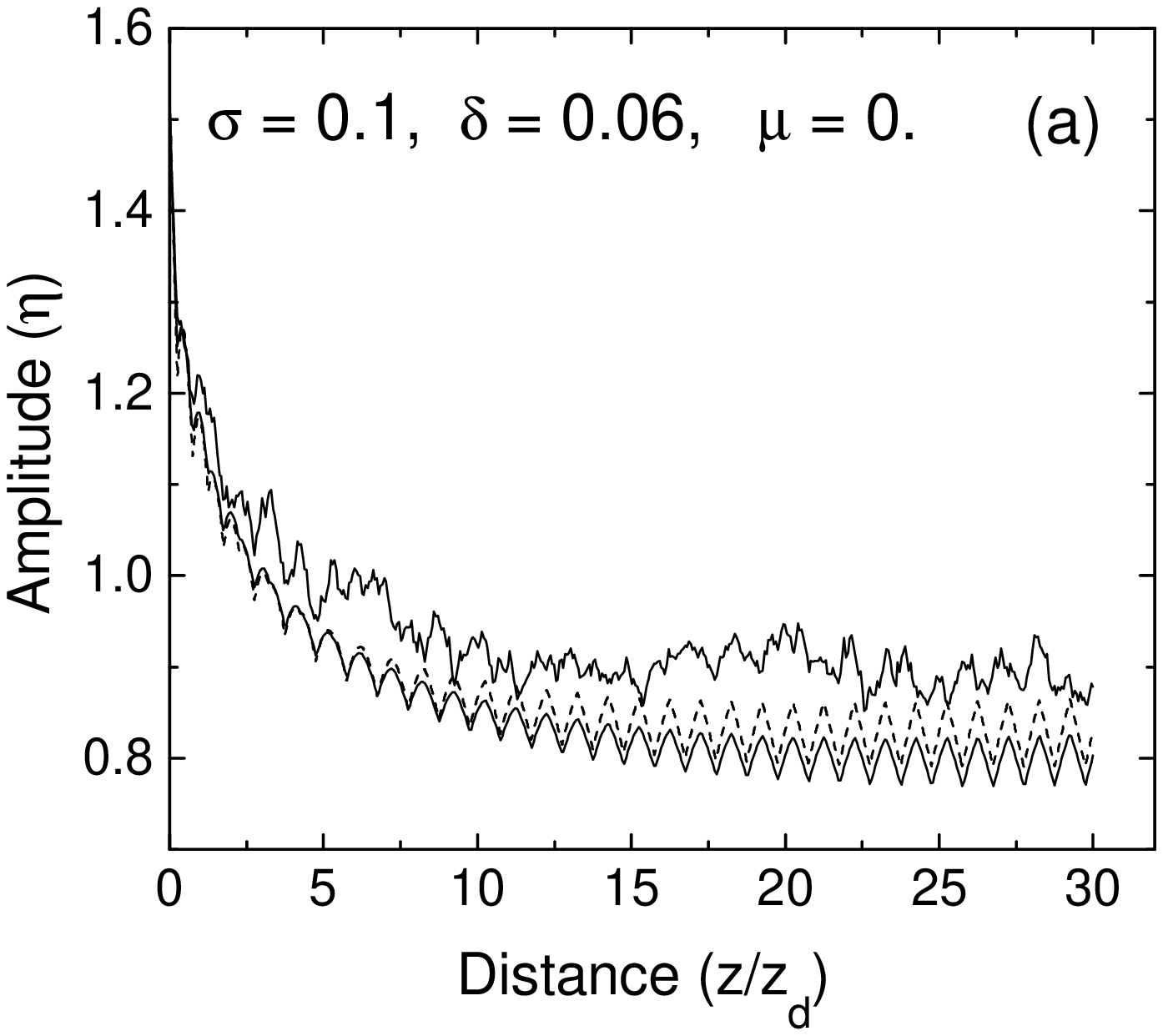}\quad
            \includegraphics[width=6cm,height=6cm,clip]{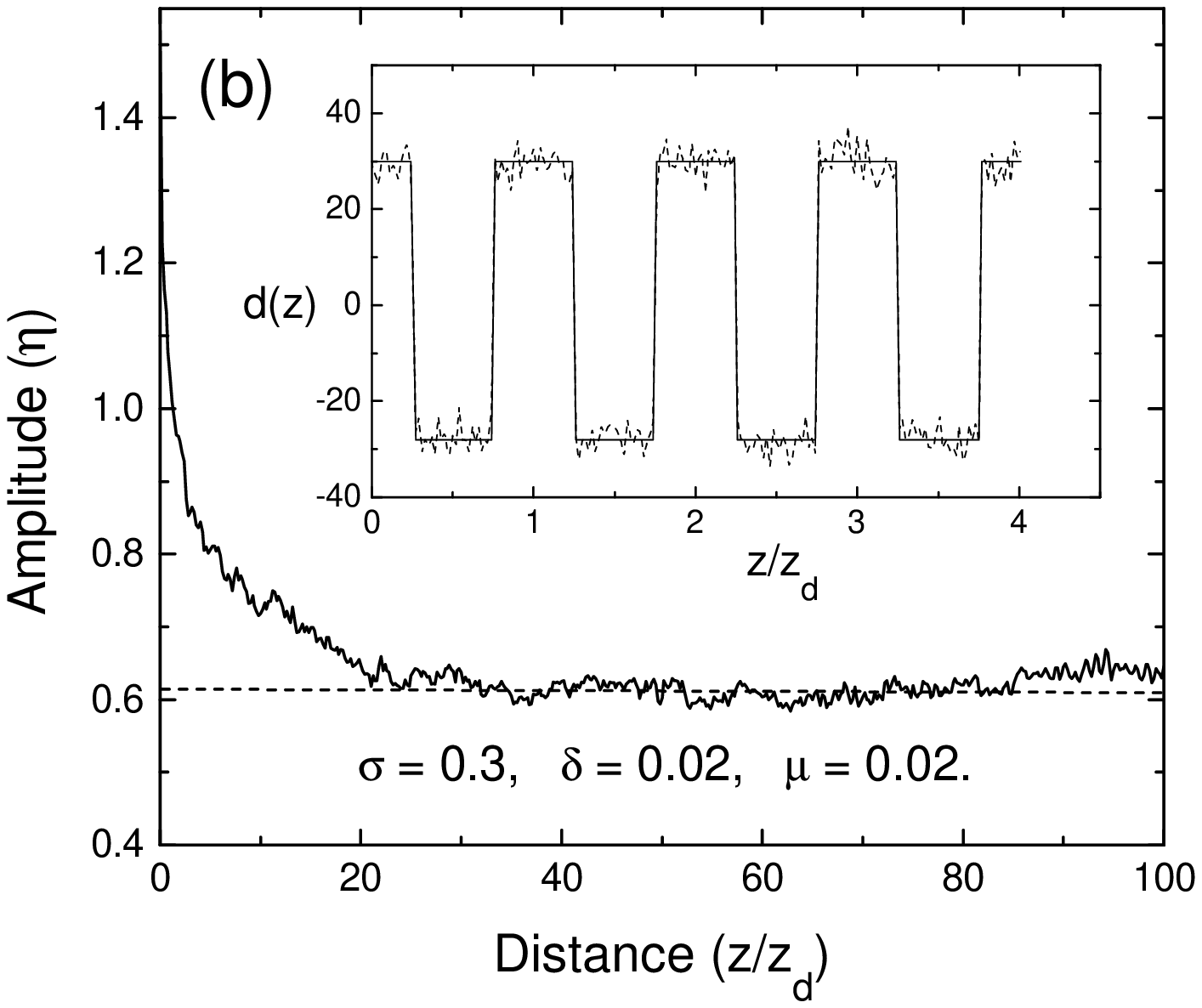}}
\caption{(a) Compensation of the pulse decay by means of linear
amplification gives rise to a stable DM soliton:  broken line -
stochastic NLS Eq.(\ref{main}), dashed line - ODE (\ref{DM-ODE}),
solid line - mean field Eq.(\ref{u4t-dm}). (b) Combined action of
linear and nonlinear amplifications on a longer distance DM
soliton propagation. The inset shows the dispersion map
(unperturbed -solid line, purturbed - dashed line). DM map
parameters and initial conditions are similar to those of figure
\ref{fig5}. }
\label{fig6}
\end{figure}

Now we try to compensate the pulse damping due to randomness of
the fiber dispersion (which is accounted in the mean field
equation as a 4th order derivative term), by linear and/or
nonlinear amplification. The Fig.\ref{fig6} demonstrates that
this is indeed achievable.

\begin{figure}[htbp]
\centerline{\includegraphics[width=6cm,height=6cm,clip]{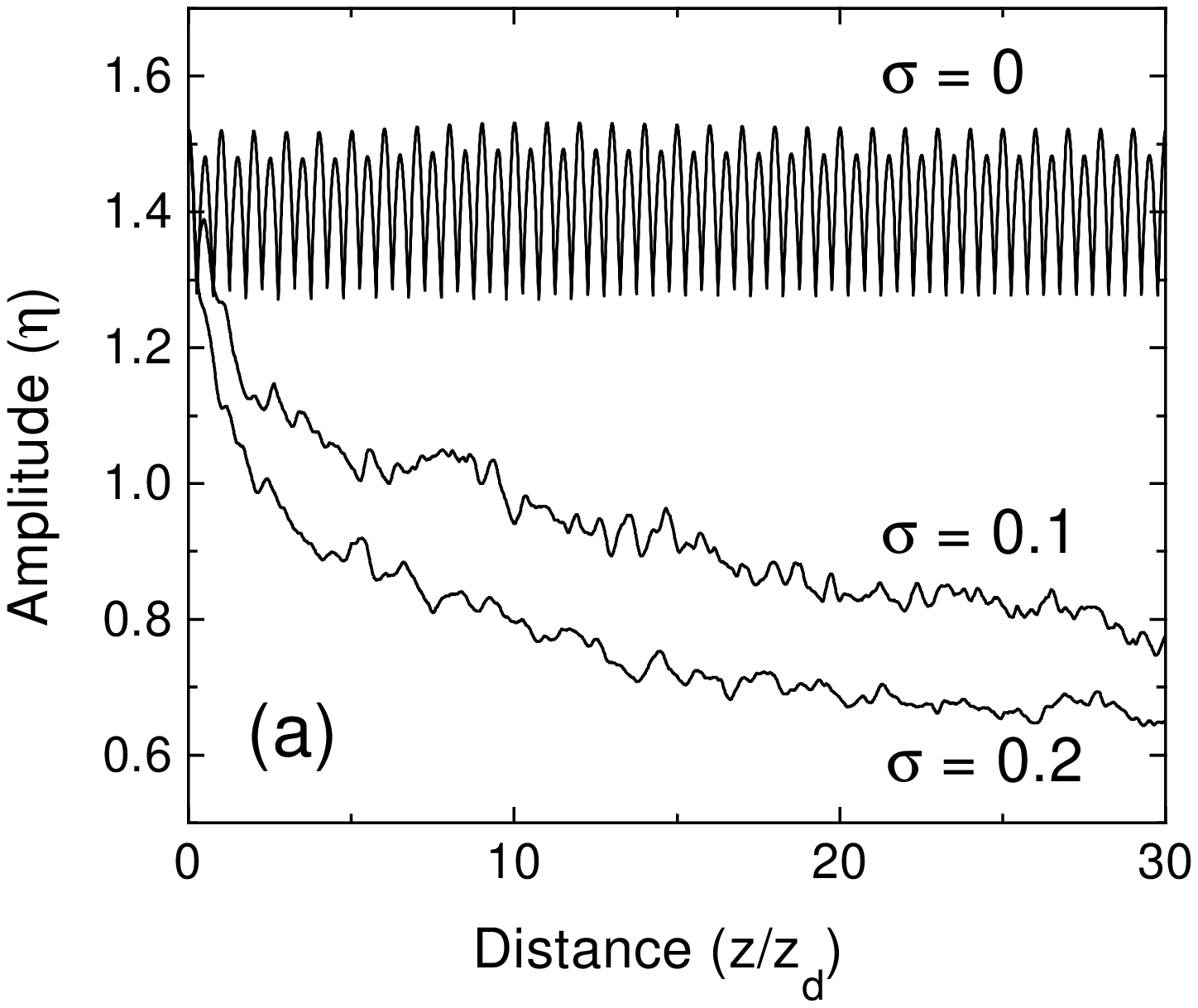}\quad
            \includegraphics[width=6cm,height=6cm,clip]{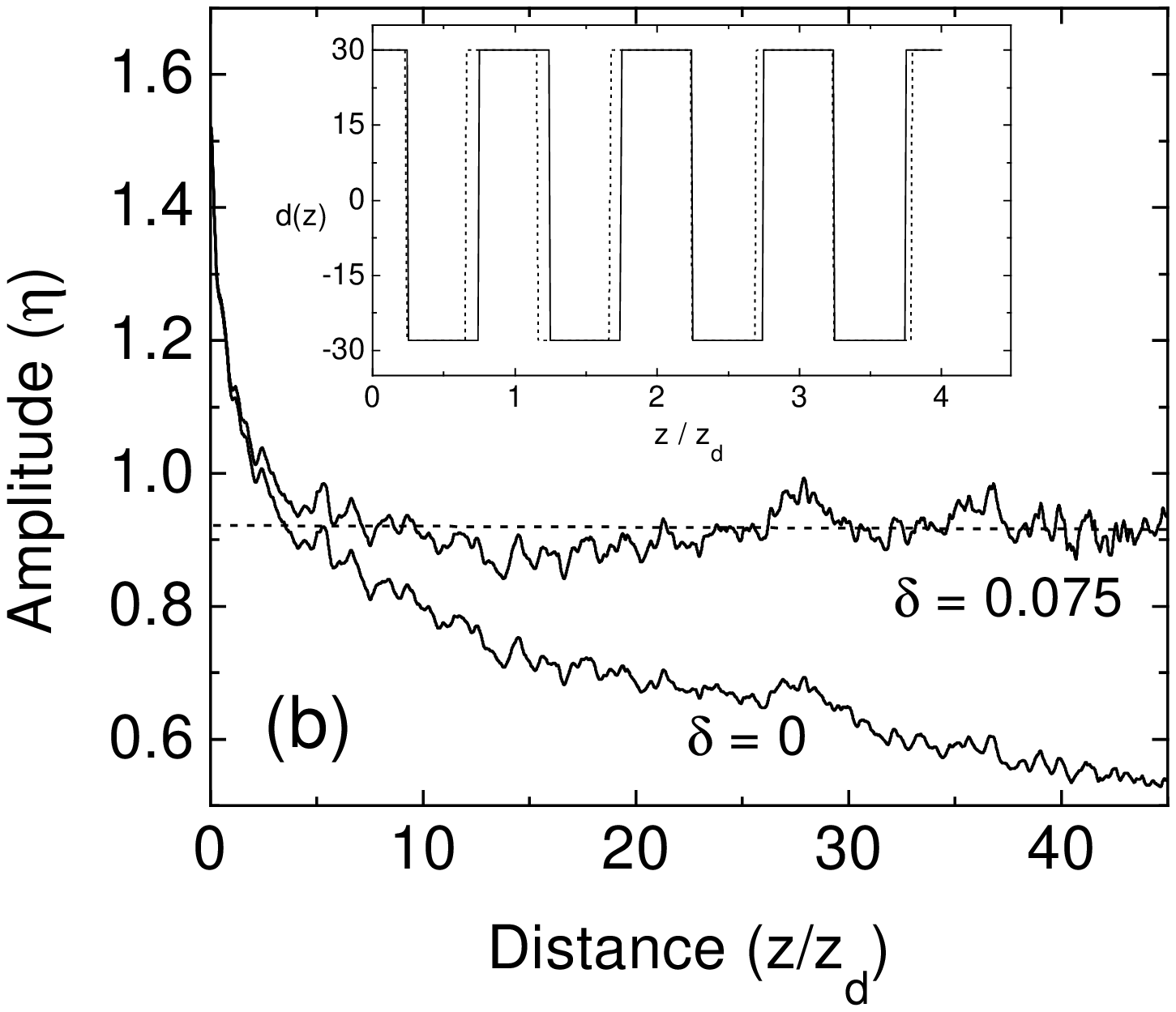}}
\caption{(a) Decay of the amplitude of a DM soliton at different
strengths ($\sigma = 0 \div 0.2$) of fluctuations of the fiber span lengths
$z_1 \cdot (1+\varepsilon(z)), \ z_2 \cdot (1+\varepsilon(z))$ according to
Eq.(\ref{main}) with $\delta = \mu = 0.$
(b) Compensation of the pulse decay (at $\sigma = 0.2$) by means of a
linear amplification ($\delta = 0.075, \ \mu = 0$) gives rise to a
stable DM soliton. The inset shows the dispersion map
(unperturbed -solid line, purturbed - dashed line). }
\label{fig7}
\end{figure}

An important kind of randomness inherent to real dispersion-managed systems
is related to the fluctuations of the lengths of fiber spans $z_1, \ z_2$.
In Fig. \ref{fig7} we illustrate that the decay of a DM soliton due to the
randomness of fiber span lengths can be compensated by a suitable linear
amplification. Comparison with the non-amplified case ($\delta = 0$) shows
the effective stabilization of the DM soliton.

Now let us analyze the fixed points of the system. These points
correspond to the stable propagation of a DM soliton under
fluctuations of the dispersion map parameters. Equating the left
hand sides of Eqs.(\ref{DM-ODE}) to zero and assuming $b \sim
\delta, \mu, \gamma \ll 1$, we obtain
\begin{eqnarray}\label{fixp}
a_0^2 &=& \frac{\sqrt{2}d_0}{A_0^2},\\ \label{A_{0}} A_0^2
&=&\frac{2\sqrt{2}\mu d_0^2}{3\gamma}\Biggl(1 \pm \sqrt{1 +
\frac{21\delta\gamma}{4 \mu^{2}d_0^2}}\Biggr),\\ \label{b_{0}}
b_0 &=& \frac{1}{2d_0}\Bigl(\frac{5\mu A_0^2}{2\sqrt{2}} -
\frac{6\gamma}{a_0^4}).
\end{eqnarray}
According to Eq.(\ref{fixp}), the stable DM soliton propagation
in optical fibers with random dispersion is possible only in the
region of anomalous dispersion ($d_{0}> 0$).

\section{Conclusion}

We have studied analytically and numerically
the propagation of optical solitons in fibers with random dispersion.
The cases when the unperturbed dispersion is uniform and periodically
modulated (DM) are considered. It is shown that the optical pulse
dynamics can be successfully described in the framework of the mean field
method. The limits of validity of the MFM is revealed by comparison with
the results of the perturbation theory, based on the IST.
It is shown that when the linear and nonlinear
gain/damping are included, corresponding mean field equation coincides with
the Swift-Hohenberg equation. Analysis revealed the existence of dissipative
solitons in fibers with fluctuating dispersion.
The system of variational equations for parameters of the DM soliton has been
derived, which takes into account the nonconservative effects due to
the randomness of the fiber dispersion. We have analyzed the fixed points of
this system and found conditions for the stable propagation of DM solitons
in a fiber with gain and fluctuating dispersion. The analytical results are
confirmed by direct numerical simulations of the full stochastic NLS equation.

\section{Acknowledgements}

This work was partially supported by the US CRDF (Award ZM-2095).
B.B. thanks the Physics Department of the University of Salerno,
Italy, for a two years research grant.

\end{document}